\documentstyle[11pt,epsfig,aaspp4]{article}
\begin{document}

\title{GRB Afterglows from Anisotropic Jets}
\author{Z. G. Dai and L. J. Gou}
\affil{Department of Astronomy, Nanjing University, 
Nanjing 210093, China \\
E-mail: daizigao@public1.ptt.js.cn (or dzg@nju.edu.cn)}

\begin{abstract}
Some progenitor models of gamma-ray bursts (GRBs) 
(e.g., collapsars) may produce anisotropic jets in which 
the energy per unit solid angle is a power-law function of 
the angle ($\propto\theta^{-k}$). We calculate light curves 
and spectra for GRB afterglows when such jets expand either 
in the interstellar medium or in the wind medium. In particular, 
we take into account two kinds of wind: one ($n\propto r^{-3/2}$) 
possibly from a typical red supergiant star and another 
($n\propto r^{-2}$) possibly from a Wolf-Rayet star. We find 
that in each type of medium, one break appears in the 
late-time afterglow light curve for small $k$ but becomes
weaker and smoother as $k$ increases. When $k\ge 2$, 
the break seems to disappear but the afterglow decays 
rapidly. Thus, one expects that the emission from 
expanding, highly anisotropic jets provides a plausible 
explanation for some rapidly fading afteglows whose 
light curves have no break. We also present good fits to 
the optical afterglow light curve of GRB 991208. Finally, 
we argue that this burst might arise from a highly 
anisotropic jet expanding in the wind ($n\propto r^{-3/2}$) 
from a red supergiant to interpret the observed 
radio-to-optical-band afterglow data (spectrum and 
light curve). 
\end{abstract}

 
\section {Introduction}

Although the study of gamma-ray bursts (GRBs) has 
been revolutionized due to observations of their 
multiwavelength afterglows and particularly due to 
determinations of their redshifts in the past few years, 
the question whether the GRB emission is spherical 
or jet-like has been unsolved. This question has 
important implications on almost all aspects of the 
GRB phenomenon, e.g., the total energy that is 
released in an explosion, the event rate, the physical 
ejection mechanism and the afterglow decay rate 
(for a review see Sari 2000). 

GRB 990123, the most energetic GRB well-studied, 
has an isotropic gamma-ray release of  $\sim 3.4 
\times 10^{54}$ ergs, which corresponds to the 
rest-mass energy of $\sim 1.9M_{\odot}$ (Kulkarni 
et al. 1999). From such an energetics, one deduced for
 the first time that the GRB emission may be highly 
collimated in this case with a half opening angle of  
$\theta_m \leq 0.2$ so that the intrinsic explosive 
energy could be reduced to $\sim 10^{51}$ -- 
$10^{52}$ ergs, which is still consistent with the 
stellar death models, e.g., collapsars (MacFadyen, 
Woosley \& Heger 2000). Another well-known feature 
of this burst is that its R-band afterglow light curve 
began to steepen at about day 2 after the burst 
(Kulkarni et al. 1999; Castro-Tirado et al. 1999; 
Fruchter et al. 1999). This steepening was first 
argued to be due to the possibility that a jet has 
evolved from the spherical-like phase to the 
sideways-expansion phase (Rhoads 1999; Sari, Piran 
\& Halpern 1999) or that we have observed the edge 
of the jet (Panaitescu \& M\'esz\'aros 1999; 
M\'esz\'aros \& Rees 1999). Alternative explanation 
for this steepening was subsequently proposed to be
that a shock expanding in a dense medium has 
undergone the transition from the relativistic phase 
to the non-relativistic phase (Dai \& Lu 1999, 2000). Since 
then, the dynamics of expanding jets has been numerically 
studied by many authors (Panaitescu \& M\'esz\'aros 1999; 
Moderski, Sikora \& Bulik 2000; Huang et al. 2000a, b;
Kumar \& Panaitescu 2000; Gou et al. 2000; Wei \& Lu 2000). 
Observationally, one marked break has also been 
observed to appear in the optical afterglow light curves 
of GRB 990510 (Harrison et al. 1999; Stanek et al. 
1999), GRB 991216 (Halpern et al. 2000; Sagar et al. 
2000a) and GRB 000301C (Rhoads \& Fruchter 2000; 
Masetti et al. 2000; Jensen et al. 2000; Sagar et al. 2000b; 
Berger et al. 2000). Holland et al. (2000) recently collected,
re-analyzed and explained all of the published photometry 
for GRB 990123 and GRB 990510. All these studies have assumed 
that both the energy per unit solid angle and the bulk Lorentz 
factor at any angle within jets are independent of the angle. 
Hereafter we refer to such jets as isotropic ones.    

In fact, both the energy per unit solid angle ($dE/d\Omega$) 
and the bulk Lorentz factor within a realistic jet may 
strongly depend on the angle $\theta$. For example, 
$dE/d\Omega\propto \theta^{-k}$ where $k\approx 3$ 
within the energetic jets expected in the collapsar model 
for GRBs of MacFadyen et al. (2000). We call such jets 
anisotropic ones. Salmonson (2000) argued that if GRBs arise 
from anisotropic jets, then the range of the viewing angle 
(viz., the angle between the line of sight and the jet axis) 
produces a range in the observed properties of  GRBs, i.e. the 
lag-luminosity relationship. In this paper, we calculate the 
emission from anisotropic jets expanding both in the interstellar 
medium (ISM) and in the wind medium. In particular, 
we consider two kinds of wind: one ($n\propto r^{-3/2}$) 
possibly from a typical red supergiant star and another 
($n\propto r^{-2}$) possibly from a Wolf-Rayet star.   
To our knowledge, this is the first numerical work about evolution 
of anisotropic jets and their emission. A preliminary analysis 
of afterglows from anisotropic jets was made by M\'esz\'aros, 
Rees \& Wijers (1998), but our results are different from their 
analytical ones because we take into account the light travel 
effects related to different sub-jets within the jet. We also carry 
out a detailed modelling of the optical afterglow flux data of 
GRB 991208 given by  Sagar et al. (2000a) and Castro-Tirado 
et al. (2000). Finally, we argue that this burst might arise from a 
highly anisotropic jet expanding in the wind from a red supergiant
to explain the observed radio-to-optical-band afterglow data. 

\section{The Model}

An energetic anisotropic jet is supposed to be ejected 
in an explosion. We assume that such a jet is adiabatic, 
i.e., the radiative energy is a negligible fraction of the total 
energy of the jet. This assumption is valid if the energy 
density of the electrons accelerated by a shock, produced 
by the interaction of the jet with its surrounding medium,
is a small fraction $\epsilon_e$ of the total energy 
density of the shocked medium or if most of the 
electrons are adiabatic, i.e., their radiative cooling 
timescale is larger than that of the jet expansion 
(Dai, Huang \& Lu 1999; B\"ottcher \& Dermer 2000). 

\subsection{Dynamics}

The central question in the next study is how the Lorentz 
factor $\gamma$ at angle $\theta$ within the jet evolves 
with the observer time $t$ because all other quantities 
that appear in the observed flux are functions of 
$\gamma$ and of the jet radius and medium density. 
To obtain $\gamma$, we assume that the kinetic energy 
per unit solid angle within the jet has the following 
distribution:
\begin{equation}
 \varepsilon(\theta)\equiv\frac{dE}{d\Omega}=\left \{
       \begin{array}{ll}
         \varepsilon_0  & {\rm if}\,\, \theta\leq\theta_0 \\
         \varepsilon_0\left(\frac{\theta}{\theta_0}\right)^{-k} 
             & {\rm if}\,\, \theta_0<\theta<\theta_m,
        \end{array}
       \right.
\end{equation}
where $\theta_m$ is the half opening angle of the jet. 
The ejected mass per unit solid angle, $M_{\rm ej}(\theta)
=M_0$, is assumed to be independent of the angle 
(at least over some range of angles). A motivation for this 
assumption is that a different distribution of the ejected 
mass affects only evolution of the jet at very early times, 
but at later times the jet expands based on the 
Blandford-McKee's (1976) self-similar solution during 
the relativistic phase and the Sedov-Taylor self-similar 
solution during the non-relativistic phase. Thus, the initial 
Lorentz factor at angle $\theta$ can be written as
\begin{equation}
 \gamma_0(\theta)=\left \{
       \begin{array}{ll}
         \frac{\varepsilon_0}{M_0c^2}+1 & {\rm if}\,\, 
         \theta\leq\theta_0 \\\left(\frac{\varepsilon_0}
        {M_0c^2}\right)\left(\frac{\theta}{\theta_0}
        \right)^{-k} +1               & {\rm if}\,\, 
        \theta_0<\theta<\theta_m.
        \end{array}
       \right.
\end{equation}
The evolution of the Lorentz factor $\gamma(\theta, r)$ 
can be calculated from the following equation
\begin{equation}
M(r)\gamma^2+M_0\gamma=M(r)+M_0+
\varepsilon(\theta)/c^2,
\end{equation}
where $M(r)$ is the mass of the swept-up medium per 
unit solid angle. Equation (3) expresses the conservation 
of energy and it applies to only the case without heating 
of the ejected material by the reverse shock. In this paper 
we neglect sideways expansion of the jet. Evidence for 
this consideration is that an observation of GRB 990123 
on 7 February 2000 by HST (Fruchter et al. 2000), 
combined with the previous observations, indicates that 
the steepening of the R-band afterglow light curve of this 
burst during a period of $\sim 378$ days is still roughly 
consistent with the edge effect of an ultra-relativistic jet of a fixed 
opening angle expanding in a homogeneous medium proposed by 
M\'esz\'aros \& Rees (1999). Further evidence is that, as argued by 
Jaunsen et al. (2000), the rapid initial decline, the sharp break in the 
optical light curve, and the spectral properties of the GRB 980519 
afterglow are best interpreted as being due to an ultra-relativistic 
jet of a fixed opening angle expanding in a wind medium 
with $n\propto r^{-2}$. 
 
The external medium density is assumed to be a power-law 
function of radius: $n(r)=Ar^{-s}$, where  $A=n_*\times 1\,
{\rm cm}^{-3}$ for an $s=0$ (homogeneous) medium (e.g., ISM), 
$A=3\times 10^{35}A_*\,{\rm cm}^{-1}$ for an $s=2$  wind 
possibly from a Wolf-Rayet star as the GRB progenitor (Chevalier 
\& Li 1999, 2000), and $A=10^{30}\bar{A}_*\,{\rm cm}^{-3/2}$ 
for an $s=3/2$ wind possibly from a typical red supergiant star 
(Fransson, Lundquist \& Chevalier 1996).  For convenience, 
$x$ is defined as the radius $r$ scaled to 
\begin{equation}
r_0=\left[\frac{(3-s)\varepsilon}{m_pc^2A\gamma_0^2}
\right]^{1/(3-s)}=\left \{
       \begin{array}{lll}
       2.7\times 10^{17}\,{\rm cm}\,\varepsilon_{53}
       ^{1/3}n_*^{-1/3}\gamma_{0,2}^{-2/3}\,\,  & {\rm if}
       \,\, s=0\\
        2.2\times 10^{16}\,{\rm cm}\,\varepsilon_{53}A_*^{-1}
             \gamma_{0,2}^{-2}\,\,  & {\rm if}\,\, s=2\\
         4.6\times 10^{14}\,{\rm cm}\,\varepsilon_{53}^{2/3}\bar{A}_*^{-2/3}
             \gamma_{0,2}^{-4/3}\,\,  & {\rm if}\,\, s=3/2,
       \end{array}
       \right.
\end{equation}
where $\varepsilon_{53}=\varepsilon(\theta)/10^{53}
{\rm ergs}$ and $\gamma_{0,2}=\gamma_0(\theta)/10^2$.  
Furthermore, we have 
\begin{equation}
M(r)=\frac{m_pn(r)r^3}{3-s},
\end{equation}
where $m_p$ is the proton mass. 

The solution of equation (3) is 
\begin{equation}
\gamma(\theta, r)=\frac{1}{2M}\left[\sqrt{M_0^2+
4M(M+M_0+\varepsilon/c^2)}-M_0\right].
\end{equation}
The dimensionless radius $x$ and the time $t'$ in the 
frame comoving with the jet evolve with the observer time 
$t$ based on
\begin{equation}
\frac{dx}{dt}=\frac{c\beta}{r_0(1-\beta\mu)},
\end{equation} 
\begin{equation}
\frac{dt'}{dt}=\frac{1}{\gamma(1-\beta\mu)},
\end{equation} 
where $\beta=\sqrt{1-1/\gamma^2}$, $\mu=\cos\theta$, 
and the line of sight has been taken to be the jet axis. 
The solution of equations (7) and (8) combined 
with (6) gives the dynamics of a sub-jet at angle $\theta$ 
within the jet. 

\subsection{Synchrotron Radiation}
  
We consider synchrotron radiation of the electrons accelerated by the 
shock. To calculate the spectrum and light curve, one needs to determine 
three crucial frequencies: the synchrotron self-absorption frequency 
($\nu_a'$), the typical synchrotron frequency ($\nu_m'$) and the cooling 
frequency ($\nu_c'$). We assume a power law distribution of the electrons 
accelerated by the shock: $dn_e/d\gamma_e\propto \gamma_e^{-p}$ for 
$\gamma_e\ge\gamma_{m}$, where $p$ is the index of 
the electron energy distribution and $\gamma_{m}=[(p-2)/
(p-1)](m_p/m_e)\epsilon_e\gamma$ is the minimum Lorentz 
factor. If $\epsilon_B$ is assumed to be the ratio of the magnetic 
to thermal energy densities of the shocked medium,  the magnetic 
field strength in the shocked medium can be approximated by  
$B'=[32\pi\epsilon_B\gamma(\gamma-1)nm_pc^2]^{1/2}$.

Under these assumptions, the typical synchrotron frequency can be written: 
$\nu_m'=\gamma_m^2eB'/(2\pi m_ec)$. The cooling frequency,
$\nu_c'=\gamma_c^2eB'/(2\pi m_ec)$,  is clearly determined 
by the Lorentz factor $\gamma_c$, at which an electron is cooling on 
the dynamical expansion timescale ($t'$, measured in the comoving frame).
According to Sari, Piran \& Narayan (1998), 
$\gamma_c=6\pi m_ec/(\sigma_TB'^2t')$ where $\sigma_T$ is the Thomson
scattering cross-section. Scaled with the typical values of the involved 
parameters, these two frequencies, measured in the comoving frame, are 
derived, 
\begin{equation}
\nu_m'=\left \{
       \begin{array}{lll}
       3.7\times 10^9\,{\rm Hz}\left(\frac{\epsilon_e}{0.1}
       \right)^2\left(\frac{\epsilon_B}{0.01}\right)^{1/2}
       \left(\frac{p-2}{p-1}\right)^2n_*^{1/2}\gamma^{1/2}
       (\gamma-1)^{5/2}\,\,  & {\rm if}\,\, s=0\\
       9.5\times 10^{10}
       \,{\rm Hz}\left(\frac{\epsilon_e}{0.1}\right)^2\left(\frac{
       \epsilon_B}{0.01}\right)^{1/2}\left(\frac{p-2}{p-1}
       \right)^2\varepsilon_{53}^{-1}A_*^{3/2}\gamma_{0,2}^2
       \gamma^{1/2}(\gamma-1)^{5/2}x^{-1}\,\,  & {\rm if}\,\, s=2\\
       3.7\times 10^{13}
       \,{\rm Hz}\left(\frac{\epsilon_e}{0.1}\right)^2\left(\frac{
       \epsilon_B}{0.01}\right)^{1/2}\left(\frac{p-2}{p-1}
       \right)^2\varepsilon_{53}^{-1/2}\bar{A}_*\gamma_{0,2}
       \gamma^{1/2}(\gamma-1)^{5/2}x^{-3/4}\,\,  & {\rm if}\,\, s=3/2;
       \end{array}
       \right.
\end{equation}
\begin{equation}
\nu_c'=\left \{
       \begin{array}{lll}
       2.8\times 10^{28}\,{\rm Hz}\left(\frac{\epsilon_B}{0.01}
       \right)^{-3/2}n_*^{-3/2}\gamma^{-3/2}(\gamma-1)^{-3/2}
       t'^{-2}\,\,  & {\rm if}\,\, s=0\\
       1.7\times 10^{24}\,{\rm Hz}\left(\frac{\epsilon_B}{0.01}
       \right)^{-3/2}\varepsilon_{53}^3A_*^{-9/2}\gamma_{0,2}
      ^{-6}\gamma^{-3/2}(\gamma-1)^{-3/2}x^3t'^{-2}\,\,  & 
     {\rm if}\,\, s=2\\
     2.8\times 10^{16}\,{\rm Hz}\left(\frac{\epsilon_B}{0.01}
       \right)^{-3/2}\varepsilon_{53}^{3/2}\bar{A}_*^{-3}\gamma_{0,2}
      ^{-3}\gamma^{-3/2}(\gamma-1)^{-3/2}x^{9/4}t'^{-2}\,\,  & 
     {\rm if}\,\, s=3/2,     
     \end{array}
       \right.
\end{equation}
where $t'$ is in units of 1 s. The synchrotron self-absorption frequency is 
usually smaller than $\nu_m'$ and $\nu_c'$ a few hours after the burst. 
In this case, the optical depth due to synchrotron self-absorption 
at $\nu_a'$ can be approximated by 
\begin{equation}
\tau_{ab}(\nu_a')\approx \frac{5}{3-s}\frac{enr}{B'\gamma_p^5}
\left(\frac{\nu_a'}{\nu_p'}\right)^{-5/3}\equiv 1,
\end{equation} 
where $\gamma_p=\min(\gamma_m, \gamma_c)$ and 
$\nu_p'=\min(\nu_m', \nu_c')$ (see, e.g., Panaitescu \& Kumar 2000). 
At early times of the afterglow, all the electrons accelerated behind 
the shock cool in a timescale smaller than the dynamical expansion 
timescale of the jet at angle $\theta$, i.e., they are in the fast cooling 
regime, implying $\nu_c'<\nu_m'$, in which case we find 
the self-absorption frequency,
\begin{equation}
\nu_a'=\left \{
       \begin{array}{lll}
       0.15\,{\rm Hz}\left(\frac{\epsilon_B}{0.01}
       \right)^{6/5}\varepsilon_{53}^{1/5}n_*^{8/5}\gamma_{0,2}^{-2/5}
       \gamma^{6/5}(\gamma-1)^{6/5}x^{3/5}t'\,\,  & {\rm if}\,\, s=0\\
       7.5\times 10^3\,{\rm Hz}\left(\frac{\epsilon_B}{0.01}
       \right)^{6/5}\varepsilon_{53}^{-3}A_*^{24/5}\gamma_{0,2}^6
      \gamma^{6/5}(\gamma-1)^{6/5}x^{-3}t'\,\,  & {\rm if}\,\, s=2\\
      1.3\times 10^{12}\,{\rm Hz}\left(\frac{\epsilon_B}{0.01}
       \right)^{6/5}\varepsilon_{53}^{-7/5}\bar{A}_*^{16/5}\gamma_{0,2}
      ^{14/5}\gamma^{6/5}(\gamma-1)^{6/5}x^{-21/10}t'\,\,  & 
     {\rm if}\,\, s=3/2.
      \end{array}
       \right.
\end{equation}
At later times, $\nu_m'>\nu_c'$, in which case the self-absorption 
frequency is given by
\begin{equation}
\nu_a'=\left \{
       \begin{array}{lll}
       4.2\times 10^8\,{\rm Hz}\left(\frac{\epsilon_e}{0.1}
       \right)^{-1}\left(\frac{\epsilon_B}{0.01}\right)^{1/5}
       \left(\frac{p-2}{p-1}\right)^{-1}\varepsilon_{53}^{1/5}n_*^{3/5}
       \gamma_{0,2}^{-2/5}\gamma^{1/5}(\gamma-1)^{-4/5}x^{3/5}
       \,\,  & {\rm if}\,\, s=0\\
      3.1\times 10^{10}
       \,{\rm Hz}\left(\frac{\epsilon_e}{0.1}\right)^{-1}\left(\frac{
       \epsilon_B}{0.01}\right)^{1/5}\left(\frac{p-2}{p-1}
       \right)^{-1}\varepsilon_{53}^{-1}A_*^{9/5}\gamma_{0,2}^2
       \gamma^{1/5}(\gamma-1)^{-4/5}x^{-1}\,\,  & {\rm if}\,\, s=2\\
      3.5\times 10^{13}
       \,{\rm Hz}\left(\frac{\epsilon_e}{0.1}\right)^{-1}\left(\frac{
       \epsilon_B}{0.01}\right)^{1/5}\left(\frac{p-2}{p-1}
       \right)^{-1}\varepsilon_{53}^{-2/5}\bar{A}_*^{6/5}\gamma_{0,2}^{4/5}
       \gamma^{1/5}(\gamma-1)^{-4/5}x^{-3/2}\,\,  & {\rm if}\,\, s=3/2.             
       \end{array}
       \right.
\end{equation}

The synchrotron peak specific luminosity of a ring 
($\theta\rightarrow \theta+d\theta$) in the comoving frame is given by
\begin{equation}
dL_m'=\left \{
       \begin{array}{lll}
       6.1\times 10^{29}\,{\rm ergs}\,\,{\rm s}^{-1}\,
       {\rm Hz}^{-1}\left(\frac{\epsilon_B}{0.01}\right)^{1/2}
       \varepsilon_{53}n_*^{1/2}\gamma_{0,2}^{-2}
       \gamma^{1/2}(\gamma-1)^{1/2}x^3\sin\theta d\theta
            \,\,  & {\rm if}\,\, s=0\\
       1.6\times 10^{31}\,{\rm ergs}\,\,{\rm s}^{-1}\,
       {\rm Hz}^{-1}\left(\frac{\epsilon_B}{0.01}\right)^{1/2}
       A_*^{3/2}\gamma^{1/2}(\gamma-1)^{1/2}
       \sin\theta d\theta\,\,  & {\rm if}\,\, s=2\\
       6.1\times 10^{33}\,{\rm ergs}\,\,{\rm s}^{-1}\,
       {\rm Hz}^{-1}\left(\frac{\epsilon_B}{0.01}\right)^{1/2}
       \varepsilon_{53}^{1/2}\bar{A}_*\gamma_{0,2}^{-1}\gamma^{1/2}
      (\gamma-1)^{1/2}x^{3/4}
       \sin\theta d\theta\,\,  & {\rm if}\,\, s=3/2.
       \end{array}
       \right.
\end{equation}
In the case of  $\nu_c'<\nu_m'$, the spectrum is (Sari et al. 1998) 
\begin{equation}
dL_{\nu'}'=\left \{
       \begin{array}{llll}
       dL_m'(\nu_a'/\nu_c')^{1/3}(\nu'/\nu_a')^2\,\, & {\rm if}\,\, 
             \nu'<\nu_a' \\
       dL_m'(\nu'/\nu_c')^{1/3}\,\, & {\rm if}\,\, 
             \nu_a'<\nu'<\nu_c' \\
       dL_m'(\nu'/\nu_c')^{-1/2}\,\, & {\rm if}\,\, 
             \nu_c'<\nu'<\nu_m' \\
       dL_m'(\nu_m'/\nu_c')^{-1/2}(\nu'/\nu_m')^{-p/2}\,\, 
             & {\rm if}\,\,\nu'>\nu_m'. 
       \end{array}
       \right.
\end{equation}
If $\nu_c'>\nu_m'$, the spectrum becomes (Sari et al. 1998) 
\begin{equation}
dL_{\nu'}'=\left \{
       \begin{array}{llll}
       dL_m'(\nu_a'/\nu_c')^{1/3}(\nu'/\nu_a')^2\,\, & {\rm if}\,\, 
             \nu'<\nu_a' \\
       dL_m'(\nu'/\nu_m')^{1/3}\,\, & {\rm if}\,\, 
             \nu_a'<\nu'<\nu_m' \\
       dL_m'(\nu'/\nu_m')^{-(p-1)/2}\,\, & {\rm if}\,\, 
             \nu_m'<\nu'<\nu_c' \\
       dL_m'(\nu_c'/\nu_m')^{-(p-1)/2}(\nu'/\nu_c')^{-p/2}\,\, 
             & {\rm if}\,\,\nu'>\nu_c'. 
       \end{array}
       \right.
\end{equation}

The observed total flux density at observed frequency 
$\nu$ is given by
\begin{equation}
F_\nu=\int^{\theta_m}_0\frac{dL'_{\nu\gamma
(1-\beta\mu)}}{4\pi D_L^2\gamma^3(1-\beta\mu)^3},
\end{equation}
where $D_L$ is the luminosity distance to the source.
For a flat universe with $H_0=65\,{\rm km}\,{\rm s}^{-1}\,
{\rm Mpc}^{-1}$, the luminosity distance to the source
$D_L=2c/H_0(1+z-\sqrt{1+z})=1.8\times 10^{28}\,{\rm cm}
[(1+z)/2]^{1/2}[(\sqrt{1+z}-1)/(\sqrt{2}-1)]$.

\section{Numerical Results}

Integrating equations (7), (8), and (17) numerically, we can 
easily obtain an afterglow light curve. Figures 1-3 exhibit different 
afterglow light curves at R-band ($\nu=4.4\times 10^{14}$ Hz) 
for different $k$ in the homogeneous ISM ($n_*=1$), in the wind 
medium with $s=2$ ($A_*=1$), and in the wind medium with $s=3/2$ 
($\bar{A}_*=0.01$), respectively. We choose the remaining 
parameters: $\varepsilon_0=10^{53}\,{\rm ergs}\,\,{\rm sr}^{-1}$, 
$M_0=5\times 10^{-4}M_\odot\,{\rm sr}^{-1}$, $\theta_0
=M_0c^2/(\varepsilon_0+M_0c^2)\approx 10^{-2}$ rad, 
$\theta_m=0.3$ rad, $p=2.5$, $\epsilon_e=0.1$, $\epsilon_B=0.01$,
and $z=1$. In each figure, the lines from top to bottom correspond to 
$k=0$, $0.5$, 1, 2 and 3 respectively. From these figures, we can see 
the following features:

\begin{enumerate}
\item At very early times, the afterglow from a jet expanding 
in the ISM brightens rapidly but the afterglow flux for the 
wind case is approximately a constant, no matter whether 
the jet is isotropic or anisotropic. This may provide a way of 
distinguishing between the environments as well as GRB 
progenitors in the near future because rapid, accurate locations 
of HETE-2 will be able to allow follow-up observations 
on very early-time X-ray and optical afterglows.  

\item For an isotropic jet ($k=0$) expanding in the ISM, 
there is a sharp break in the light curve at later times. The time 
at which this break occurs is close to the analytical result, 
\begin{equation}
t_{\rm jet}\approx 2\times 10^6(\varepsilon_{53}/n_*)
^{1/3}(\theta_m/0.3)^{8/3}\,\,{\rm s}, 
\end{equation}
at which the Lorentz factor of the jet is equal to the inverse of 
its half opening angle (M\'esz\'aros \& Rees 1999). But for an 
isotropic jet expanding in the wind, the break of the light curve 
is  weaker and smoother. 

\item In each type of medium, the break of the light curve 
becomes weaker and smoother as  $k$ increases. When 
$k=2$ or 3, the break seems to disappear but the afterglow 
decays rapidly. This result shows that the emission from 
expanding, highly anisotropic jets may provide an 
explanation for some rapidly fading afteglows whose light 
curves have no break (e.g., GRB 991208; see the next 
section). We note that in the preliminary analysis of 
M\'esz\'aros et al. (1998), an afterglow decays as one single 
power-law for any $k$. The reason for this difference will
be discussed in the final section.

\item At very late times, the slopes of the light curves are 
approximately equal. We also find that the slopes become 
steeper with the electron distribution index $p$.     
\end{enumerate}

In Figures 4-6 we present the observed radiation spectra computed 
for a sequence of time for a jet with $k=3$ in three kinds of medium 
for the same parameters as in Figures 1-3. It can be seen 
that even for such an anisotropic jet considered here, the observed 
radiation spectra can still be divided into four parts, which can be well 
described from low-energy to high-energy bands by power-law 
functions $F_\nu\propto \nu^\beta$, with $\beta=2$, $1/3$, 
$-(p-1)/2=-0.75$ and $-p/2=-1.25$ at $t\ge 10^4$\,s, respectively. 
Two adjacent spectrum portions are still joined very sharply. 
These results may be due to that the observed radiation spectra 
are mainly contributed by the $\theta<\theta_0$ part of the jet.

\section{The Afterglow of GRB 991208}

GRB 991208 was a long burst with a duration of 
$\sim 60$ s and a fluence of $\sim 10^{-4}\,{\rm erg}
\,{\rm cm}^{-2}$ ($>25$ keV) (Hurley et al. 1999). 
Its redshift was measured as $z=0.7055\pm 0.0005$ 
(Djorgovski et al. 1999), and thus its isotropic energy in 
$\gamma$-rays is estimated: $E_{\rm iso}\sim 1.3\times 
10^{53}$ ergs. Because its afterglows at optical, millimeter 
and radio wavelengths were detected and unusually bright 
(Sagar et al. 2000a; Castro-Tirado et al. 2000; Galama et al. 
2000), a detailed study of this burst may provide new clues 
regarding the origin of the GRB phenomenon. 

Sagar et al. (2000a) gave the observed spectral and temporal 
indexes of the optical afterglow $\beta_{\rm ob}=-0.75\pm 0.03$ 
and $\alpha_{\rm ob}=-2.2\pm 0.1$ while Galama et al. 
(2000) presented well-sampled spectra and light curves 
between radio and millimeter wavelengths for a two-week 
period, and obtained, for the first time, the evolution of the 
synchrotron self-absorption frequency $\nu_a\propto 
t^{-0.15\pm 0.12}$, the peak frequency $\nu_m\propto 
t^{-1.7\pm 0.4}$, and the peak flux density $F_m\propto 
t^{-0.47\pm 0.11}$.  The existence of one single power law 
decay for the R-band afterglow implies that the afterglow 
might be produced by one of the following models: 
({\em i}) an isotropic jet with sideways expansion, ({\em ii}) 
a relativistic fireball expanding in the ISM, and ({\em iii}) 
a relativistic fireball expanding in the wind. Galama et al. 
(2000) argued that model ({\em i}) can explain the 
observations, but models ({\em ii}) and ({\em iii}) are ruled 
out from $\alpha_{\rm ob}$ and $\beta_{\rm ob}$.  

If $p=2.5$, inferred from $\beta_{\rm ob}=-(p-1)/2
=-0.75\pm 0.03$, Figures 7 and 8 provide two good fits to the 
observed R-band afterglow ligh curve based on our jet model 
with $k=3$ in cases of media (ISM and wind with $s=2$), 
respectively. This shows that GRB 991208 might arise from 
a highly anisotropic jet. However, we note that these two cases 
may not provide further fits to the observed radiation spectra
because the evolution of the theoretical synchrotron 
self-absorption frequency, peak frequency and peak flux density 
is inconsistent with the observations by Galama et al. (2000).      

Now we assume a generic wind case: 
$n\propto r^{-s}$. In this case, we have derived the synchrotron 
self-absorption frequency $\nu_a\propto t^{-3s/[5(4-s)]}$ 
if $\nu_a<\nu_m$, and the peak frequency $\nu_m\propto 
t^{-3/2}$, and the peak flux density $F_m\propto t^{-s/(8-2s)}$ 
in the ultra-relativistic phase (Dai \& Lu 1998). 
If $s=3/2$, we easily find $\nu_a\propto t^{-0.36}$, 
$\nu_m\propto t^{-1.5}$, and $F_m\propto t^{-0.3}$. 
These scaling laws are in approximate accord with those obtained 
from the observations. This preliminary result prompts us to 
re-consider a highly anisotropic jet expanding in an $s=3/2$ wind to 
fit the afterglow data of GRB 991208. In Figure 9 we first show  
a satisfactory fit to the observed R-band light curve.    
It is interesting to note that such a wind has been inferred 
in the circumstellar medium of SN 1993J (whose progenitor 
is a red supergiant) by Fransson et al. (1996), and is possibly 
caused by a variation of the mass-loss rate from the progenitor 
or by a non-spherical geometry. In Figure 10 we further present 
a fit to the radio-to-optical-band afterglow spectrum observed 
on 1999 December 15.5 UT for the same parameters as in Figure 9.                       
Therefore, we can conclude that GRB 991208 might 
come from a highly anisotropic jet expanding in the 
wind environment ($s=3/2$) from a red supergiant.

\section{Discussion and Conclusions}      

The model presented in section 2 clearly includes the following  
assumptions. First, the initial opening angle of a highly anisotropic jet 
(e.g., $k=3$), $\theta_m$, was taken to be $0.3$, and the observer's 
angle ($\theta_{\rm obs}$) between the axis line of the jet and the line 
of sight to be zero. Because $dL_m\propto \varepsilon(\theta)$ in the 
ultra-relativistic limit, the emission from the jet mainly arises from the 
shock-accelerated electrons moving along the axis line. Thus, the flux 
density of the emission is weakly dependent of $\theta_m$, but strongly 
depends on $\theta_{\rm obs}$. For example, the flux density at 
$\theta_{\rm obs}=0.3$ is about four orders of magnitude smaller than
that of the emission at $\theta_{\rm obs}=0$. GRB 980425 was likely  
such an off-axis burst surrounded by an $s=2$ wind medium (Nakamura 1999) 
because its gamma-ray luminosity of GRB 980425 was also approximately 
four orders of magnitude smaller than that of GRB  970228/980326 
and its X-ray afterglow slowly declined.
Second, we considered jets of a fixed opening angle and neglected 
the effect of sideways expansion based on two observational facts.
As shown analytically by Rhoads (1999) and Sari, Piran \& Halpern (1999),
this effect can lead to a sharp break in an afterglow light curve.
However, numerical studies by many authors (e.g, Panaitescu \& 
M\'esz\'aros 1999; Moderski, Sikora \& Bulik 2000; Huang et al. 2000a, b;
Kumar \& Panaitescu 2000; Gou et al. 2000; Wei \& Lu 2000) show that 
the actual break, when two effects such as the equal-time surface 
and detailed dynamics of the jet are considered, is much weaker and 
smoother than that predicted analytically. 
Finally, we used the adiabatic solution for kinematics of the jet. 
This argument is based on the radiative efficiency of the jet, 
$f=\epsilon_e[t_{\rm syn}^{\prime -1}/(t_{\rm syn}
^{\prime -1}+t_{\rm exp}^{\prime -1})]$, defined by Dai et al. (1999).
At very early times, the cooling timescale due to synchrotron radiation, 
$t'_{\rm syn}$, may be much less than the expansion timescale, 
$t'_{\rm exp}$, and $f\approx 0.1(\epsilon_e/0.1)$; but at late times, 
$t'_{\rm syn}\gg t'_{\rm exp}$ and thus $f\approx 0.1(\epsilon_e/0.1)
(t'_{\rm exp}/t'_{\rm syn})\ll 1$. Therefore, it can be seen that 
the energy losses due to synchrotron radiation are insignificant 
during the whole evolution stage of the jet.
   
Our numerical results  show that for an isotropic jet ($k=0$) expanding 
in the ISM, there is a sharp break in the light curve at late times, but for 
an isotropic jet expanding in the wind, the break of the light curve is  
weaker and smoother.  This result is easily understood in terms of 
the analytical model of Dai \& Lu (1998) and Chevalier \& Li (1999, 2000): 
at early times of the jet evolution, $\gamma>\theta_m^{-1}$, in which 
stage the jet is spherical-like and thus the decay index of the afterglow 
$\alpha_1=-[s/(8-2s)+3(p-1)/4]$ in the slow cooling regime; but at later  
times, $\gamma<\theta_m^{-1}$,  the decay index becomes 
$\alpha_2=-[(6-s)/(8-2s)+3(p-1)/4]$ due to the edge effect. These indexes 
imply the steepening of the light curve $\Delta\alpha\equiv\alpha_1-
\alpha_2=(3-s)/(4-s)=0.5$ for $s=2$, 0.6 for $s=3/2$, or 0.75 for $s=0$. 
Therefore, the break in the light curve of the afterglow from a jet expanding 
in a wind is weaker and smoother than that for a homogeneous medium case.
Our results also show that for an anisotropic jet, one break seems to 
appear in the light curve at small $k$ but becomes weaker and 
smoother as  $k$ increases. This conclusion is different from that 
presented by M\'esz\'aros et al. (1998), who found that an afterglow 
decays as one single power-law for any $k$. The reason for this difference 
is that we took into account the light travel effects 
related to different sub-jets within the jet but M\'esz\'aros 
et al. (1998) didn't. It is easily understood that an anisotropic jet with 
a small $k$ can be treated as a quasi-isotropic jet with a half 
opening angle $\theta_m$, and thus the temporal decay of the flux density 
of radiation from such a jet will begin to expedite at $t=t_{\rm jet}$, 
which is given by equation (18) for the ISM case. However, for a highly 
anisotropic jet (e.g., $k=2$ or 3), the contribution of radiation from 
$\theta>\theta_0$ is in fact much smaller than that from $\theta<\theta_0$ 
due to the fact that the energy per unit solid angle ($dE/d\Omega$) 
within this jet decreases rapidly with increasing $\theta$, and thus, 
the light curve of the resulting radiation should begin to steepen 
at a very early time, $t\approx 230(\varepsilon_{53}/n_*)^{1/3}
(\theta_0/10^{-2})^{8/3}\,{\rm s}$, which is also estimated from 
equation (18) in the ISM case but for $\theta_m$ being replaced by 
$\theta_0$. Therefore, one cannot see a broken light curve at late 
times for such a highly anisotropic jet.    
  
In summary, the energy per unit solid angle within a realistic jet 
may be a power-law function of the angle ($\propto\theta^{-k}$). 
Such anisotropic jets are expected in the collapsar model of 
MacFadyen et al. (2000). We numerically calculated light curves 
and spectra of the emission from anisotropic jets expanding either 
in the interstellar medium (ISM) or in the wind medium. We also 
took into account two kinds of wind: one ($n\propto r^{-3/2}$) 
possibly from a typical red supergiant and another 
($n\propto r^{-2}$) possibly from a Wolf-Rayet star.   
Two of the main results of this work are that ({\em i}) at very 
early times, the afterglow from a jet expanding in the ISM 
brightens rapidly but the afterglow flux for the wind case is 
approximately a constant, no matter whether the jet is 
isotropic or anisotropic. Based on this conclusion, future 
observations led by HETE-2 on very early-time optical afterglows 
may be able to distinguish between the environments as well as 
GRB progenitors. ({\em ii}) In each type of medium, 
one break appears in the late-time afterglow light curve for 
small $k$, but becomes weaker and smoother as $k$ increases. 
When $k\ge 2$, the break seems to disappear but the afterglow 
decays rapidly. Therefore, one expects that the emission from 
expanding, highly anisotropic jets provides an explanation for 
some rapidly fading afteglows whose light curves have no break. 
We also presented good fits to the optical afterglow flux data of 
GRB 991208. Finally, to interpret the observed 
radio-to-optical-band afterglow data of this burst, 
we argued that it might arise from a highly anisotropic jet 
expanding in the wind ($n\propto r^{-3/2}$) from a red supergiant. 

\acknowledgments
 
We would like to thank the referee for several valuable comments 
that allowed us to improve the manuscript, and D. M. Wei for reading 
carefully the manuscript. This work was supported by the National 
Natural Science Foundation of China (grant 19825109) and 
the National 973 Project.

\clearpage
\begin{figure}
\begin{picture}(100,250)
\put(0,0){\includegraphics{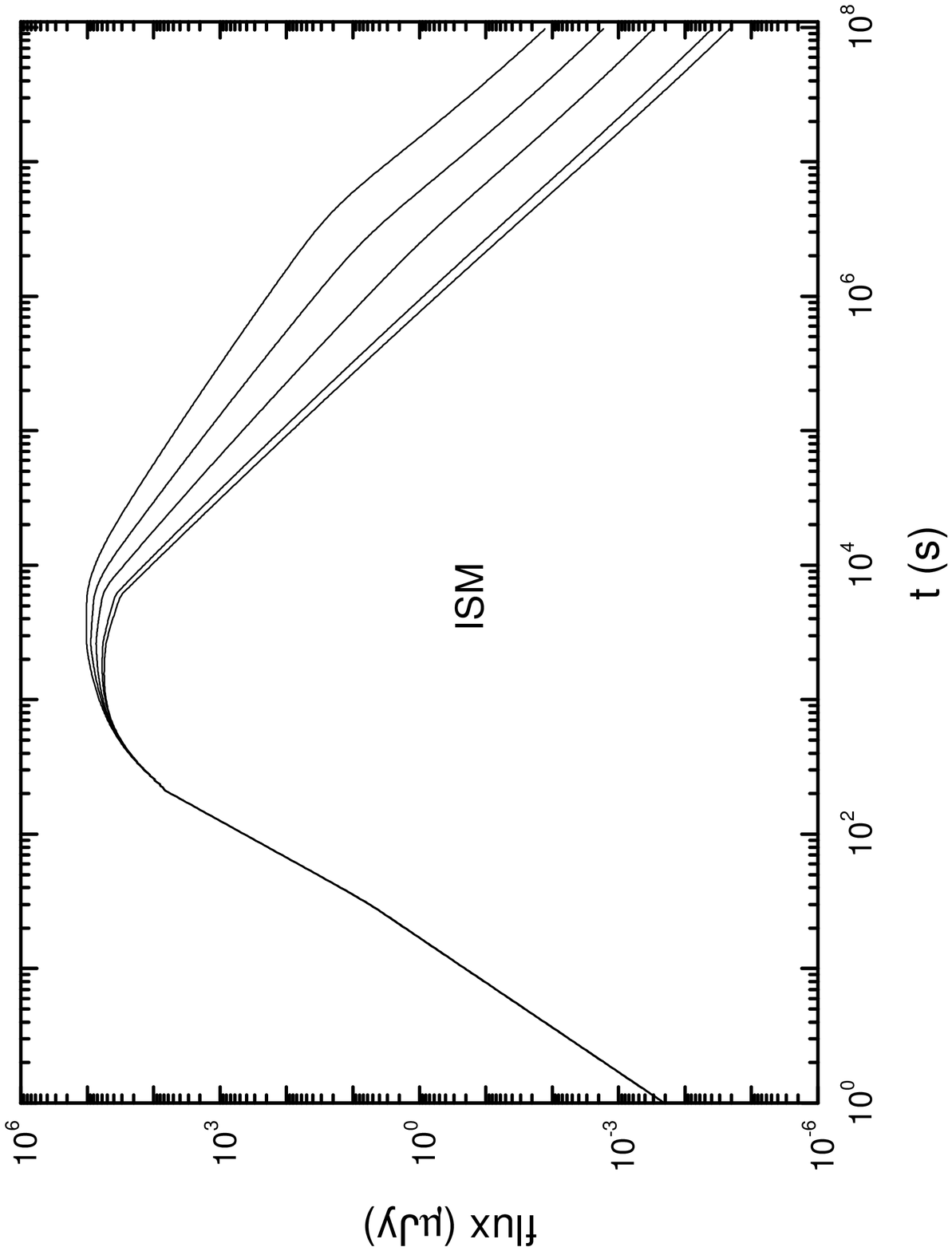}}
\end{picture}
\caption
{Light curves of R-band afterglows from anisotropic jets expanding 
in the ISM ($s=0$). The lines from top to bottom correspond  to $k=0$, $0.5$, 
1, 2 and 3 respectively.  The other parameters for the model are seen in the text.}
\end{figure}

\clearpage
\begin{figure}
\begin{picture}(100,250)
\put(0,0){\includegraphics{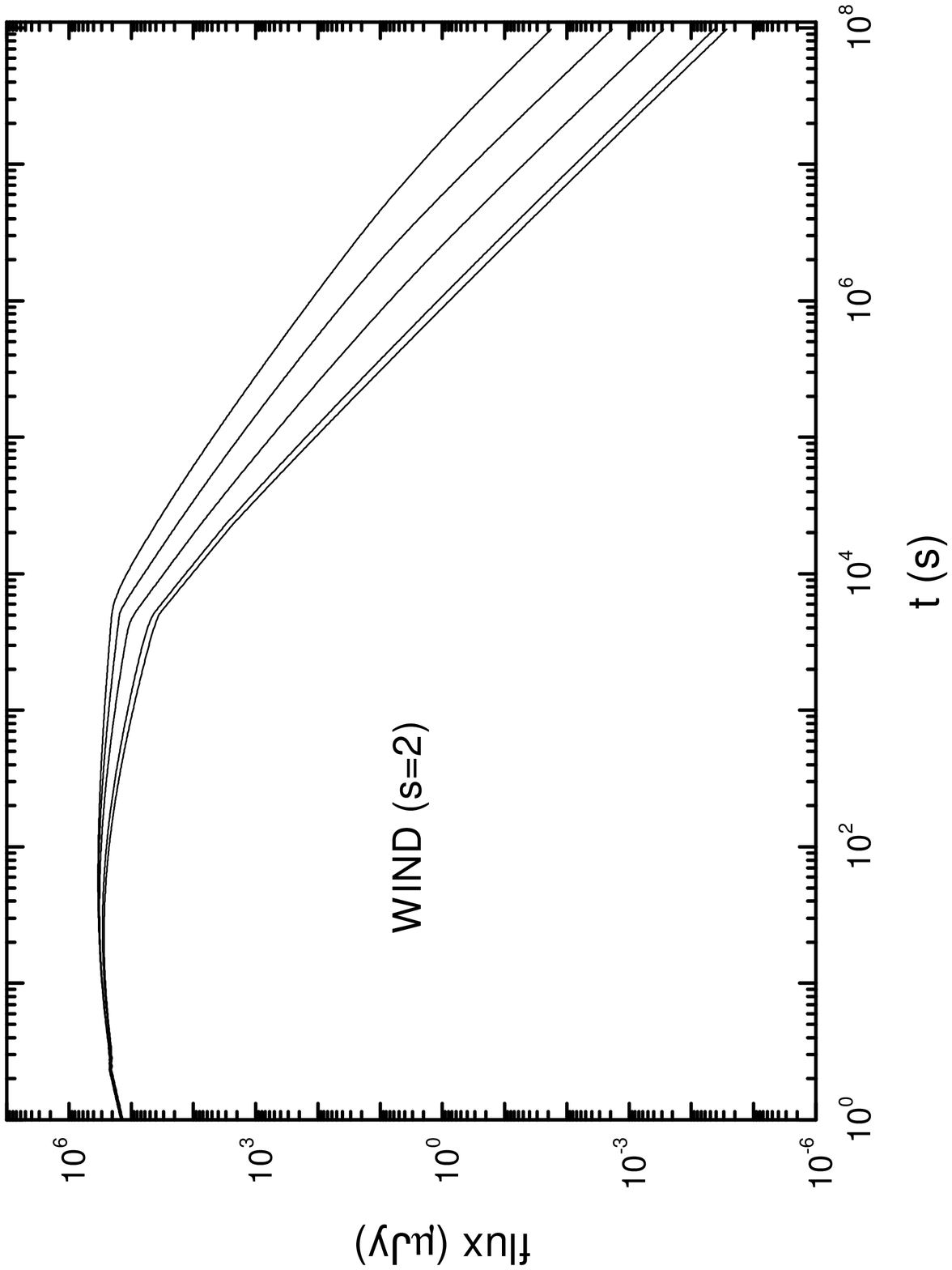}}
\end{picture}
\caption
{The same as in Figure 1 but for the $s=2$ wind.}  
\end{figure}

\clearpage
\begin{figure}
\begin{picture}(100,250)
\put(0,0){\includegraphics{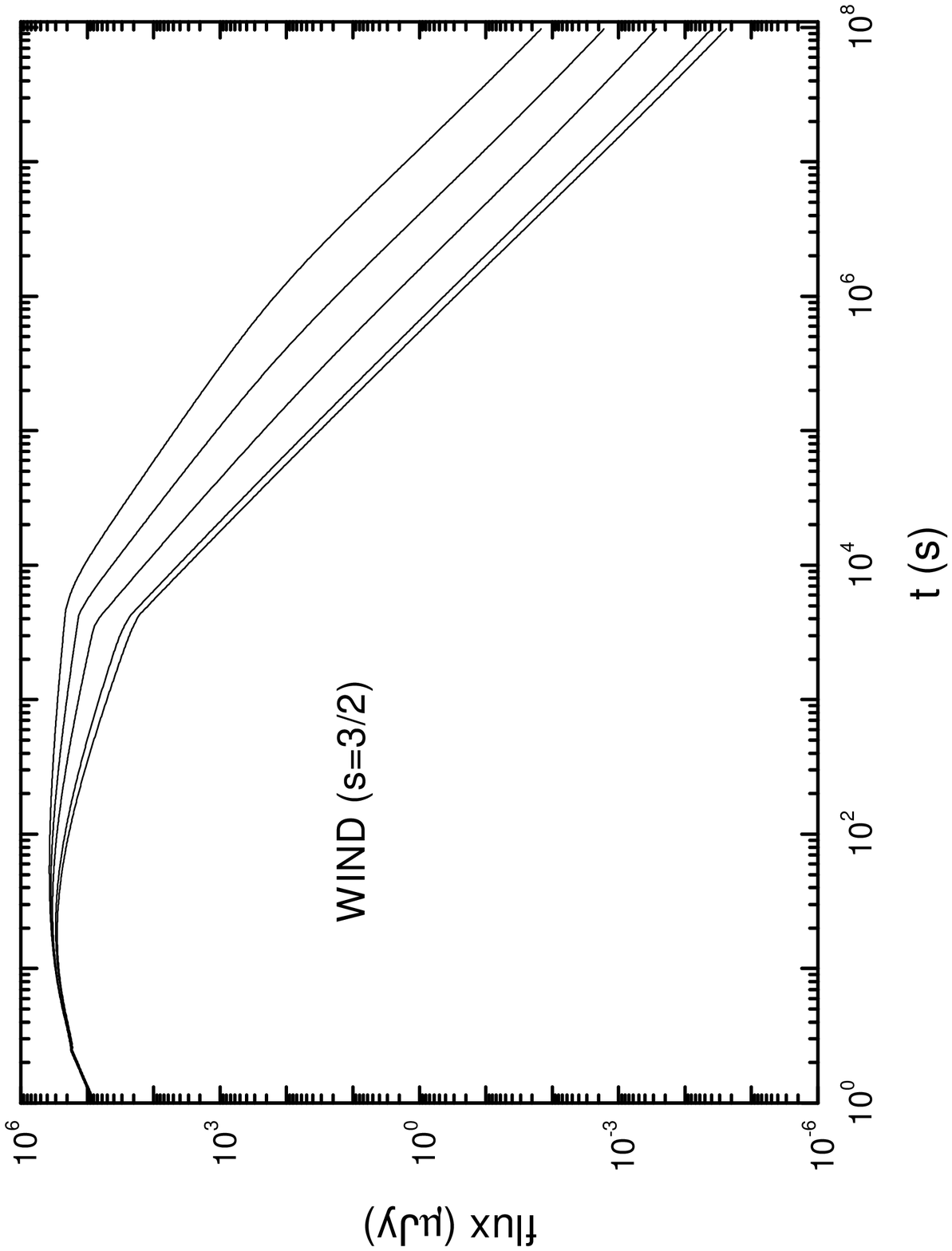}}
\end{picture}
\caption
{The same as in Figure 1 but for the $s=3/2$ wind.}  
\end{figure}

\clearpage
\begin{figure}
\begin{picture}(100,250)
\put(0,0){\includegraphics{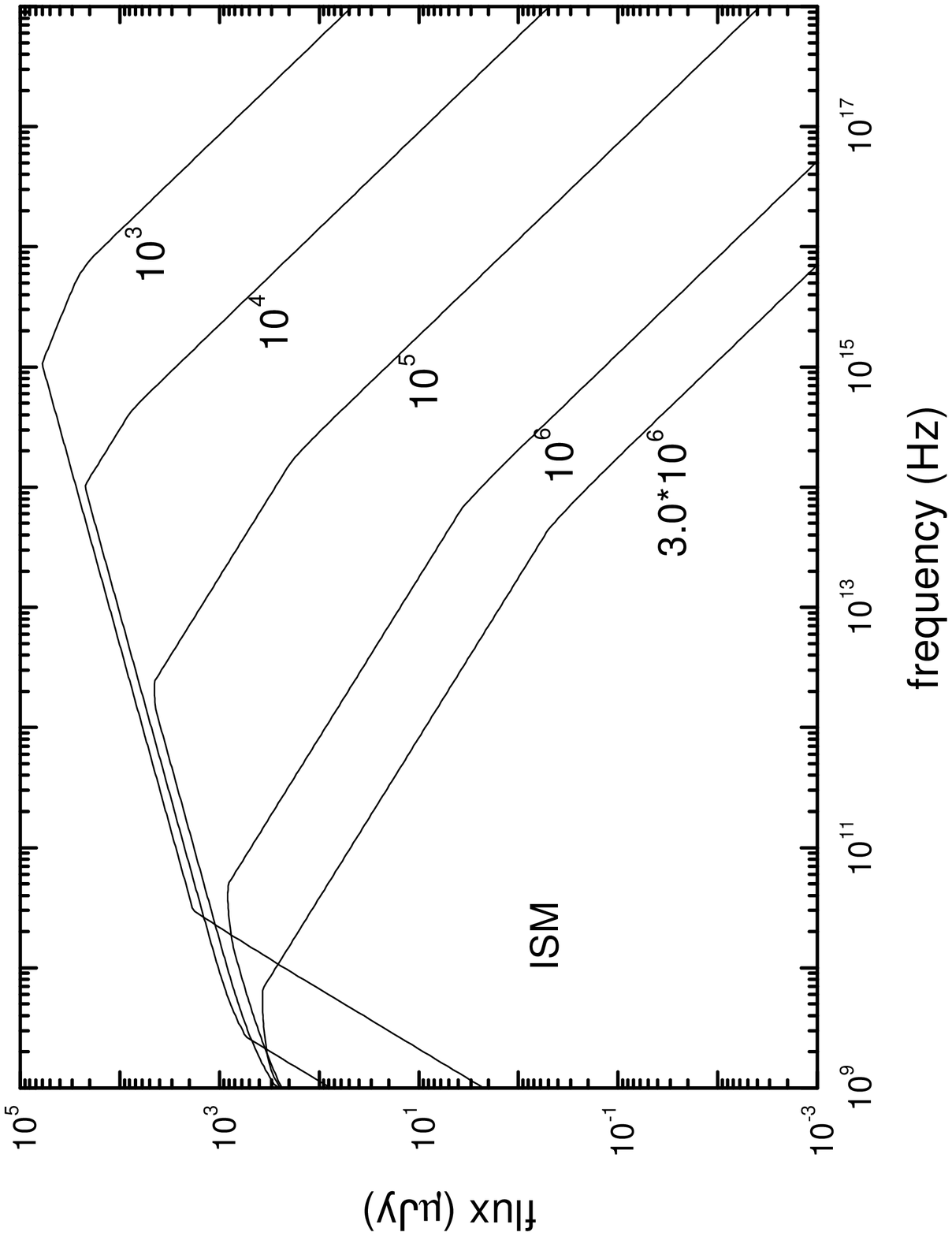}}
\end{picture}
\caption
{Evolution of the observed radiation spectra for a $k=3$ jet expanding 
in the ISM ($s=0$). The numbers mark the observed times 
in units of second. The other parameters for the model are the same as 
in Figure 1.}
\end{figure}

\clearpage
\begin{figure}
\begin{picture}(100,250)
\put(0,0){\includegraphics{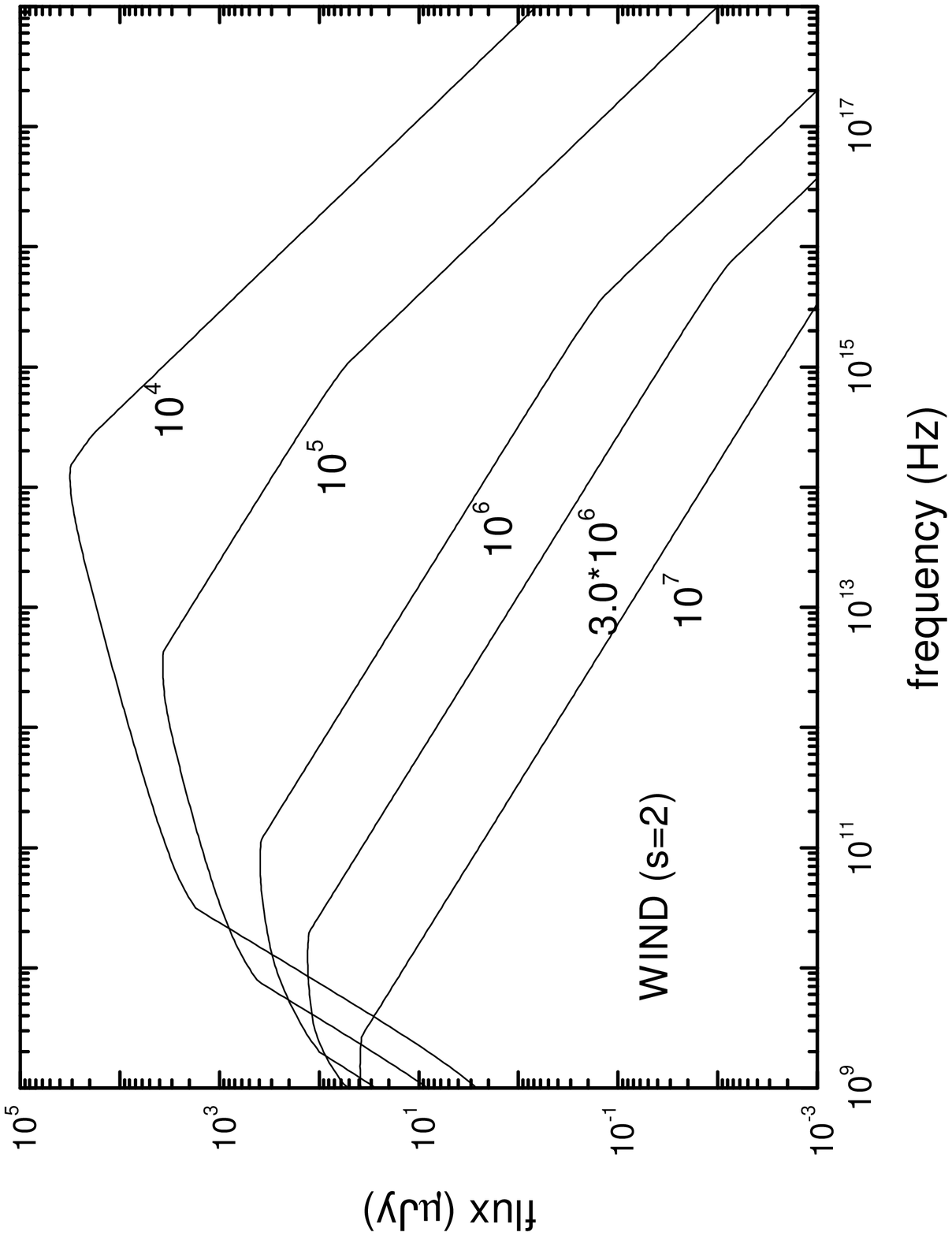}}
\end{picture}
\caption
{The same as in Figure 4 but for the $s=2$ wind.}  
\end{figure}

\clearpage
\begin{figure}
\begin{picture}(100,250)
\put(0,0){\includegraphics{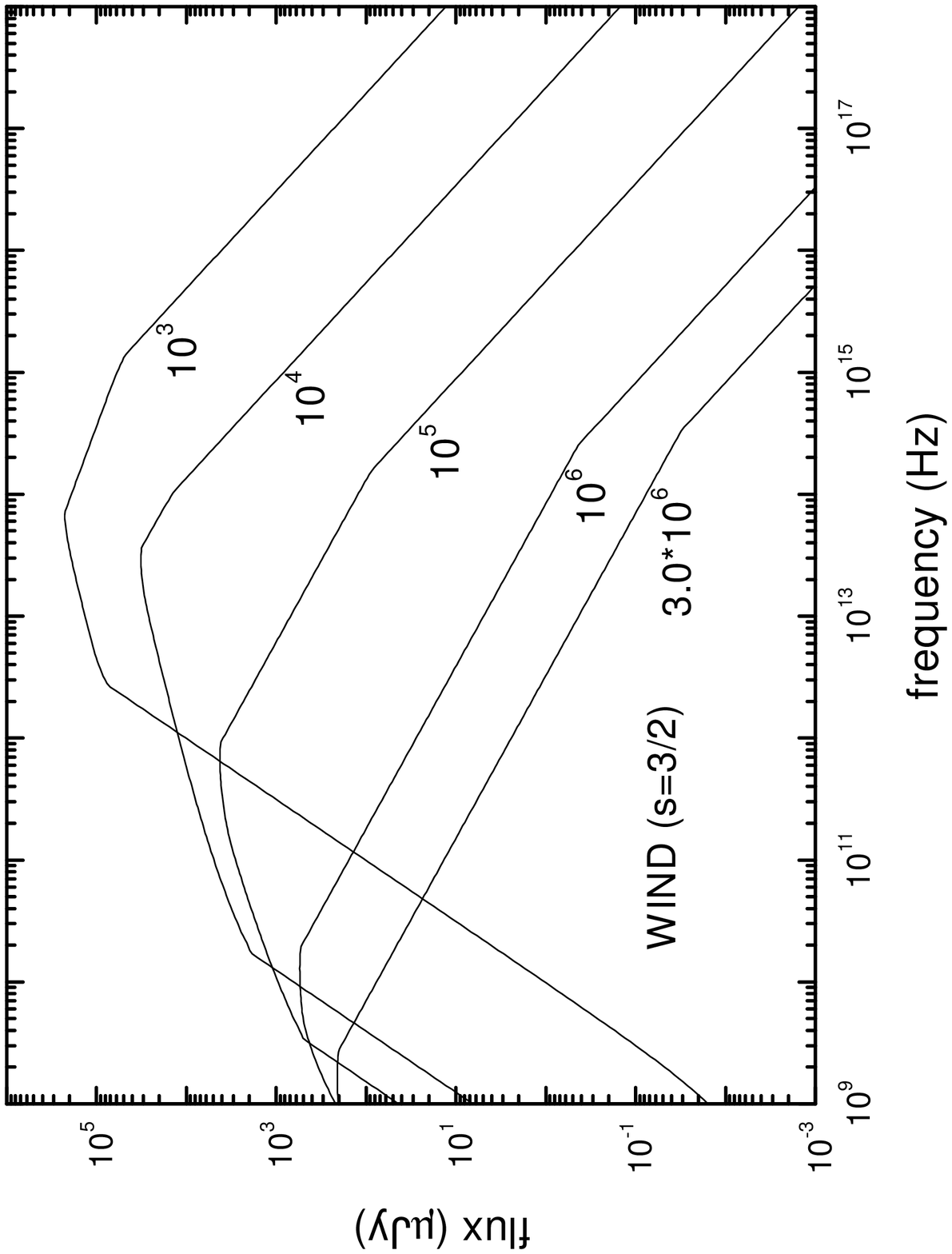}}
\end{picture}
\caption
{The same as in Figure 4 but for the $s=3/2$ wind.}  
\end{figure}

\clearpage
\begin{figure}
\begin{picture}(100,250)
\put(0,0){\includegraphics{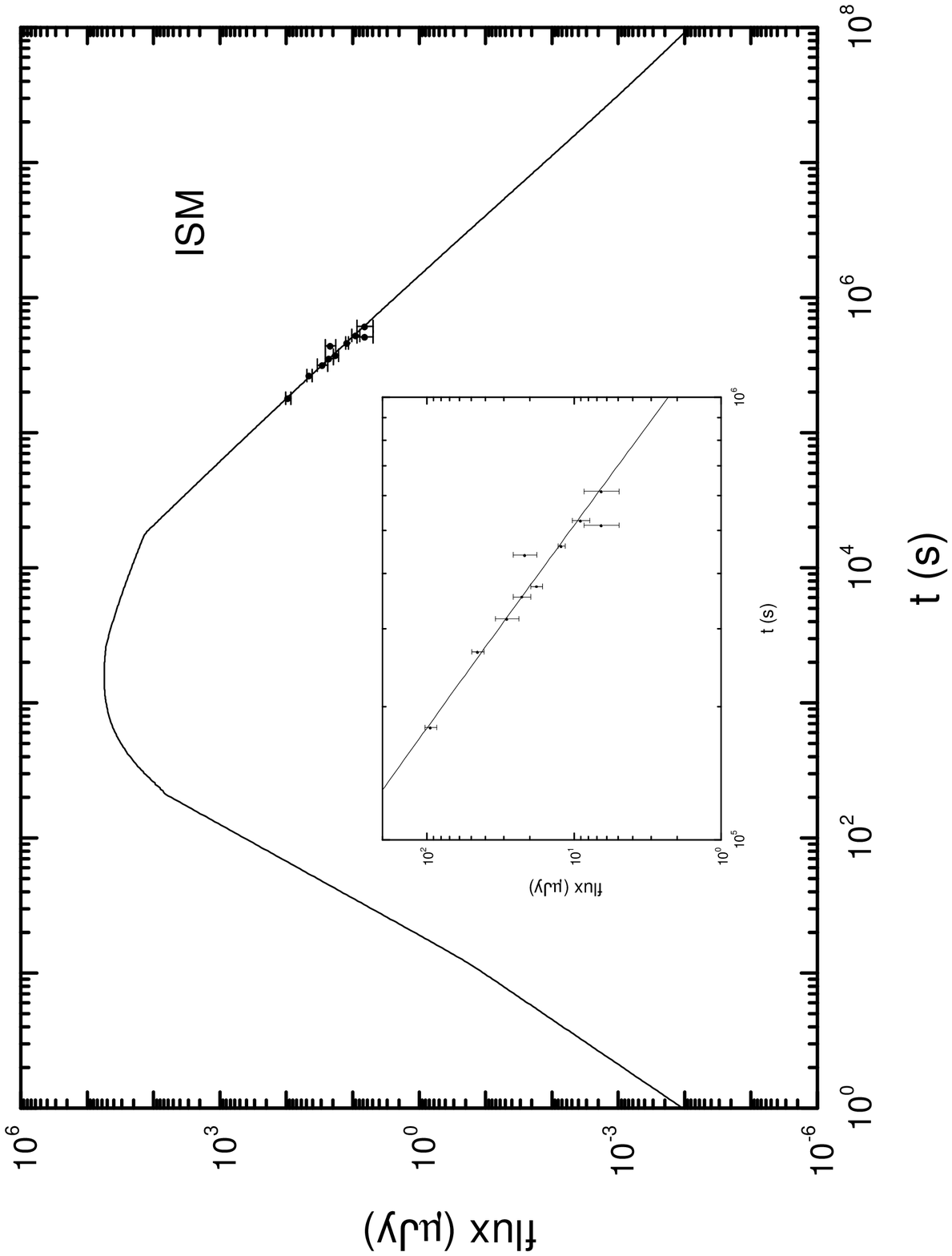}}
\end{picture}
\caption
{Comparison between the observed and theoretically calculated 
light curves for the R-band afterglow of GRB 991208. The data 
are taken from Sagar et al. (2000a) and Castro-Tirado et al. (2000), 
and the model light curve is calculated for a $k=3$ jet expanding 
in the ISM ($s=0$) when an observer is located on the jet axis. 
The model parameters are chosen: $\varepsilon_0=
10^{53}\,{\rm ergs}\,\,{\rm sr}^{-1}$, $M_0=5\times 10^{-4}
M_\odot\,{\rm sr}^{-1}$, $\theta_0=10^{-2}$ rad, $\theta_m
=0.3$ rad, $p=2.5$, $n_*=1.0$, $\epsilon_e=0.25$ and 
$\epsilon_B=0.01$.  The insert shows a clearer fit.}
\end{figure}

\clearpage
\begin{figure}
\begin{picture}(100,250)
\put(0,0){\includegraphics{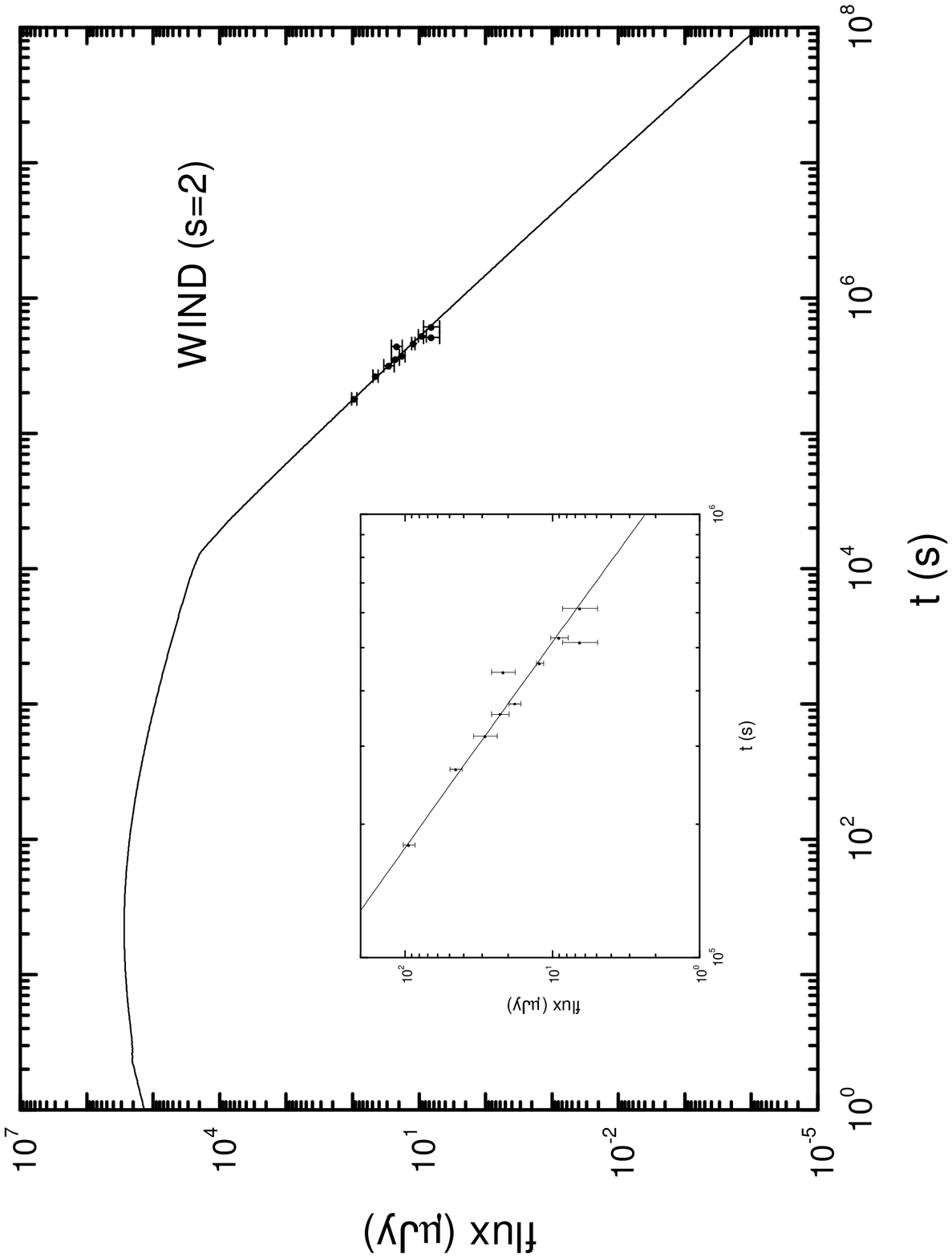}}
\end{picture}
\caption
{The same as in Figure 7 but for $s=2$ (wind), $A_*=1.0$ 
and $\epsilon_e=0.21$.}
\end{figure}

\clearpage
\begin{figure}
\begin{picture}(100,250)
\put(0,0){\includegraphics{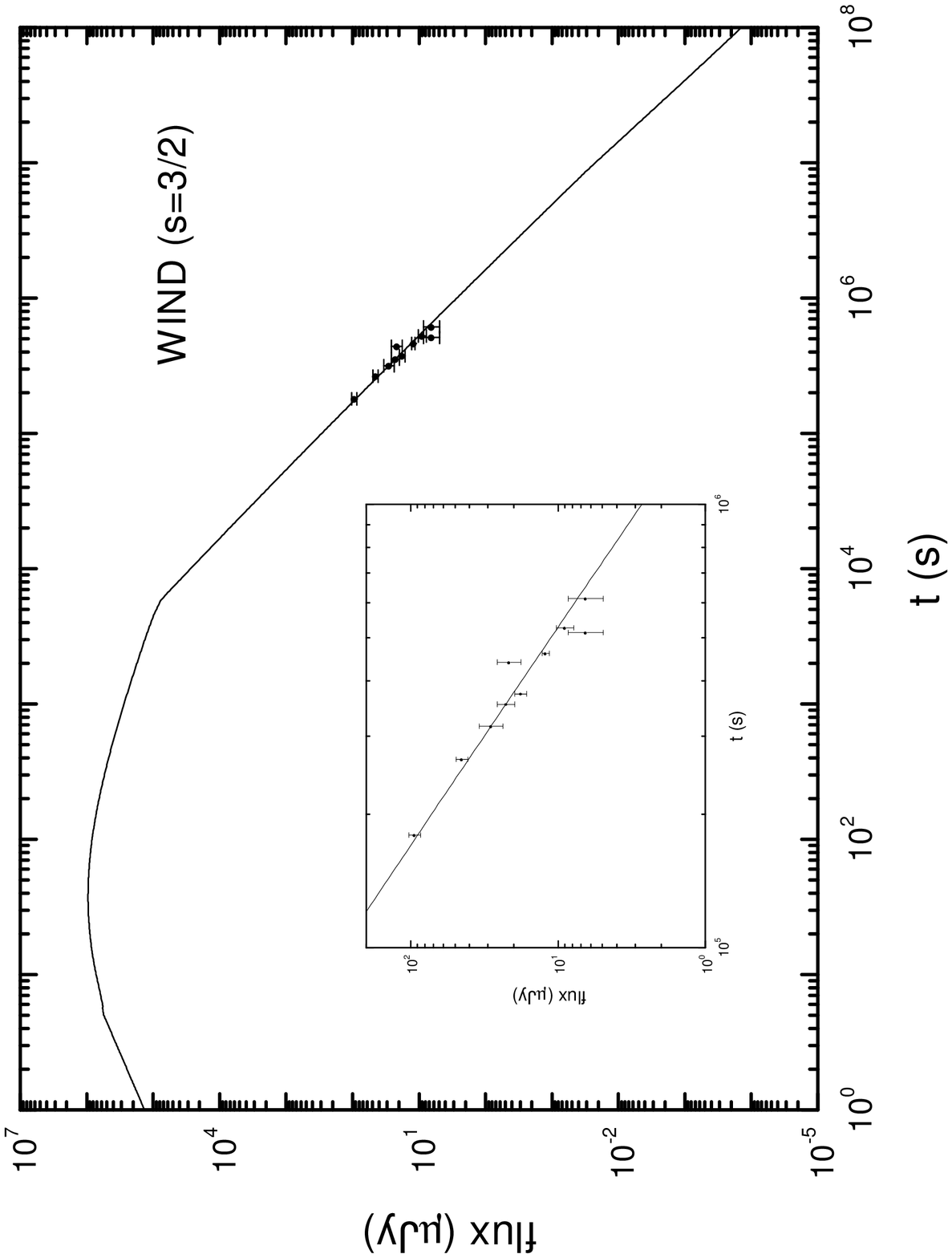}}
\end{picture}
\caption
{The same as in Figure 7 but for the $s=3/2$ wind case and for the 
following model parameters: $\varepsilon_0=1.5\times 10^{53}\,
{\rm ergs}\,\,{\rm sr}^{-1}$, $\theta_0=0.01$ rad, $\bar{A}_*=8\times 10^{-3}$, 
$\epsilon_e=0.16$, $\epsilon_B=10^{-3}$, $p=2.5$, $k=3$, and 
$\theta_m=0.3$ rad.}
\end{figure}

\clearpage
\begin{figure}
\begin{picture}(100,250)
\put(0,0){\includegraphics{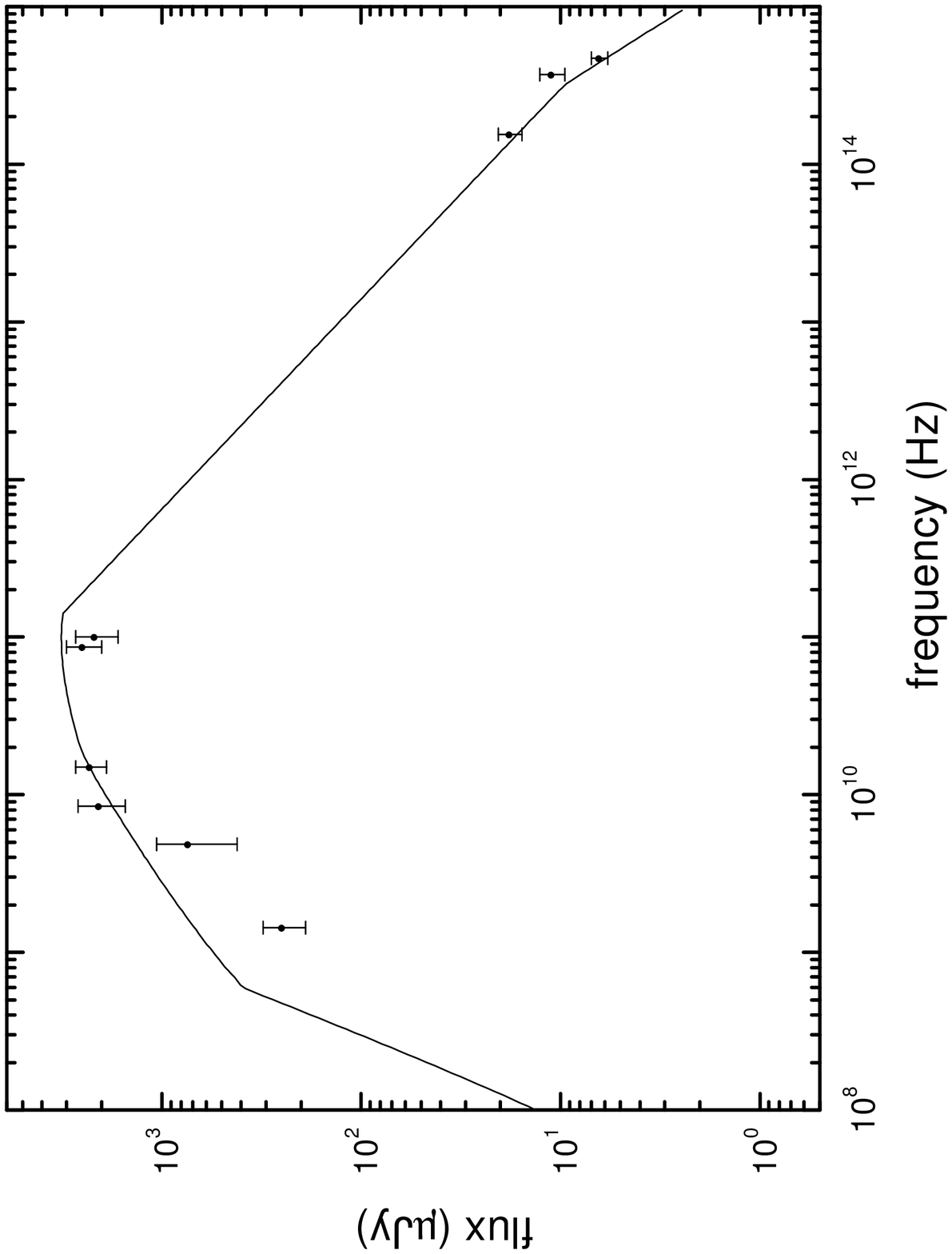}}
\end{picture}
\caption
{Comparison between the observed and theoretically calculated 
spectra of the GRB 991208 afterglow on 1999 December 15.5 UT. 
The data are taken from Galama et al. (2000), Sagar et al. (2000a) 
and Castro-Tirado et al. (2000). The model and its parameters are 
the same as in Figure 9.}
\end{figure} 

\end{document}